\documentstyle[12pt]{article}
\begin{document}
\textheight=22 true cm
\textwidth=15 true cm
\normalbaselineskip=24 true pt
\normalbaselines
\topmargin -0.5 true in
\bibliographystyle{unsrt}
\def\sl{\em}
\def\vslash{v\!\!\!/}
\def\vpslash{{v^\prime}\!\!\!\!/}
\def\lslash{l\!\!\!/}
\def\epslash{\epsilon\!\!\!/}
\def\be{\begin{equation}}
\def\ee{\end{equation}}
\def\bea{\begin{eqnarray}}
\def\eea{\end{eqnarray}}
\newcommand{\gtlt}{\begin{array}{c}>\\[-10pt]<\end{array}}

%PAPER BEGINS HERE
\setcounter{page}{0}
\thispagestyle{empty}

\begin{flushright}
{\sf SINP-TNP/96-04}\\
{\sf TIFR/TH/96-13}\\
{\sf March 1996}
\end{flushright}
\begin{center}
{\Large \bf Applicability of  Heavy Quark Effective Theory  \\
to  the Radiative  Decay ${\unboldmath B\to K^* \gamma}$}\\[5mm]
{\large\sf Debrupa Chakraverty, Triptesh De and Binayak Dutta-Roy}\\
{\sl Theory Group, Saha Institute of Nuclear Physics,\\
1/AF Bidhannagar, Calcutta - 700 064, India}\\[5mm]
and\\[5mm]
{\large\sf Anirban Kundu \footnote{Electronic address:
akundu@theory.tifr.res.in}}\\ 
{\sl  Theoretical
Physics Group,\\
Tata Institute of Fundamental Research,\\
Homi Bhabha Road, Bombay - 400 005, 
India}\\[5mm] 
\end{center}

\begin{abstract}

In the context of the observed decay $B \rightarrow K^* \gamma$  the 
applicability of the Heavy Quark Effective Theory (HQET), treating both 
b and s as heavy quarks, is examined.  We show that the heavy s-quark 
approximation, as can be found in the literature, is not reliable.  
\end{abstract}
\bigskip

\newpage
\centerline {\Large \bf I. INTRODUCTION}
 The  observed \cite{pdg} rare decay $B\to K^* \gamma $
(proceeding, as it does, through a 
 flavor changing neutral current and absent at the tree level
 in the Standard Model(SM)),  has
attracted considerable attention, as it enables the testing of loop effects 
 \cite{bertolini} involving
 the CKM matrix elements $V_{tb}$ and $V_{ts}$; furthermore 
 additional contributions in the loop
 stemming from new bosons and fermions present in most of the extensions of
 the SM,  
holds forth the possibility that
 this process  provides a window to new physics effects.
  However, the usual problem of relating 
 hadron properties to quark-level
 theoretical inputs has to be addressed. While this, in principle, involves 
 the notorious and as yet rather intractable features of non-perturbative 
 QCD, approaches invoking effective Lagrangians could provide valuable 
 estimates. The Heavy Quark Effective Theory (HQET) \cite{neu} is expected to be 
 useful in this regard in so far as  the b quark is concerned. However, 
 the s quark in the final hadron can neither be considered heavy
 enough to enable the use of HQET nor sufficiently light to permit 
 the exploitation of the chiral perturbation theory in an unambiguous manner.
  Nevertheless, attempts \cite{ali} have been made to apply HQET to both b and s 
 quarks. The object of this paper is to examine the reliability of such an 
approach in the context of the decay in question, particularly
because of the physics that lie in the loop and the fact that
 it is a rare decay  which
has been observed.

The paper is organised as follows: the next section sets forth the basic 
formalism.  
In section III we analyse the results using 
the Isgur, Scora, Grinstein and Wise
(ISGW) model \cite{isgw} (modified by Amundson \cite{iisgw}) and
the Bauer, Stech and 
 Wirbel (BSW) model \cite{bsw}   for the extraction of corrections
 of order 
$\Lambda_{QCD}/m_s$ to results based 
 on the consideration that both b and s quarks are heavy, in order to arrive
 at an assessment as to whether such an approach is reliable.
Section IV  concludes the paper.

\bigskip 

\centerline{\Large \bf II. FORMALISM }

  The exclusive decay $B
\rightarrow K^* \gamma$ is expected to be well described  by the quark
level process $b \rightarrow s \gamma$ with reasonably small corrections
 from hadronic effects unlike what obtains for the
decays of the K-meson, where nonperturbative long distance QCD contributions 
are expected to be substantial. 
Accordingly, integrating out the top-quark and W-boson
fields, one arrives at an effective Hamiltonian for $B$-meson decays
which comprises of a sum of ten operators \cite{grin, grig}.     
\begin {equation} 
H_{eff} = -{4 G_F \over \sqrt 2} V_{tb} V_{ts}^*
\sum_{i=1}^{10} C_i(\mu) O_i(\mu)
\end{equation}
where $C_i(\mu)$, the Wilson coefficients, arise from the
renormalisation group equations to provide the scaling down
 to a subtraction point  \footnote{Whether
the subtraction point is to be taken at $\mu=m_b$ is a point of
debate. The popular range is from $m_b/2$ to $2m_b$.}
appropriate to the problem viz. $\mu \approx m_b$.  Of the ten
operators $O_i$ the one that contributes to $B
\rightarrow K^* \gamma$ is 
\begin{equation}
O_7 = {e \over 32 \pi^2} F_{\mu \nu} \big[ m_b {\bar s} \sigma^{\mu
\nu} (1 + \gamma_5) b + m_s {\bar s} \sigma^{\mu \nu} (1 - \gamma_5)b
\big]
\end{equation}
where $\sigma^{\mu \nu} = {i\over 2} [ \gamma^\mu , \gamma^\nu ]$ and
$F_{\mu \nu}$ is the electromagnetic field tensor. 
The relevant Wilson coefficient \cite{grin} is given by
\begin{eqnarray}
C_7 (\mu) & = & {[{\alpha_s (M_W)\over \alpha_s(m_b)}]^{16/23}}\{C_7(M_W)
-{8 \over 3}C_8(M_W)[1-{({\alpha_s(m_b)\over \alpha_s(M_W)})^{2/23}}]
\nonumber\\
& & +{232\over 513}[1-{({\alpha_s(m_b)\over \alpha_s(M_W)})^{19/23}}]\}
\end{eqnarray}
with 
\begin{equation}
C_7(m_W)=-{x\over 2}[{{2\over 3}x^2+{5\over 12}x-{7\over 12}
\over  {(x-1)^3}}
 -{({3\over 2}x^2 -x)\ln x\over {(x-1)^4}}]
\end{equation}
and
\begin{equation}
C_8(m_W)=-{x\over 4}[{{{1\over 2}x^2-{5\over 2}x-1}\over {(x-1)^3}}
+{3x\ln x\over {(x-1)^4}}]
\end{equation}
where $x=m_t^2/m_W^2$, and $C_8$  is the  Wilson
coefficient accompanying the operator $O_8$, which is 
\begin{equation}
O_8={\alpha\over {4\pi}}({\bar s}\gamma_\mu{1\over 2}(1-\gamma_5)b)
({\bar e}\gamma_\mu\gamma_5e)
\end{equation}
Even though one may later treat both b and s as heavy in the context
of strong interaction effects, nevertheless taking cognisance of
the fact that $m_b >> m_s$, only the term involving
$m_b$ in the operator $O_7$ need be retained. Thus the
 matrix element of interest  becomes
\begin{equation}
\langle K^* \gamma \vert O_7 \vert \bar B \rangle 
= { e \over 32 \pi^2} q_\mu \eta_\nu 
\langle K^* \vert \bar s [\gamma^\mu , \gamma^\nu ] (1 + \gamma_5) b
\vert \bar B \rangle m_b
\end{equation}
where $q_\mu $ is the four momentum of the photon and $\eta_\mu$ its
polarisation vector.
It should be noted that here the matrix element of a tensor current between
hadronic states is involved for which not much information is available.
 Heavy quark effective theory (HQET), however,
comes to the rescue as the associated heavy-quark spin 
symmetry enables one to express the matrix element of the tensor operator
 in terms of the vector
and axial vector form factors, which  also occur in 
semi-leptonic decays and  may be 
estimated in different phenomenological models.
This is brought to bear upon the problem by first noticing that one
may write the four-momentum carried away by the photon  $q =
p-p^\prime$	($p$ and $p^\prime$ being the four-momenta of the initial
and final mesons) as  $q = M_B v -M_{K^*}v^\prime$ where $v$ and
$v^\prime$ are the four-velocities of the initial and final mesons
respectively. Furthermore, if both the bottom and the strange-quarks
are taken to be heavy, the four-velocities of these quarks may,
in the leading order,  be approximated to be
 that of the corresponding mesons, and
accordingly,  the ensuing relations
\begin {equation}
\vslash ~~b ~~= ~~b
\end{equation}
\begin{equation}
\vpslash ~~s ~~= ~~s
\end{equation}
permits one to write 
\begin {eqnarray}
\langle K^* \gamma \vert O_7 \vert \bar B \rangle
& = & { e \over 16 \pi^2 } m_b\eta_{\mu}
\big[ M_{K^*} \langle K^* \vert {\bar s}
\gamma^\mu (1+\gamma
_5)b \vert \bar B \rangle\nonumber\\
 & & +M_B \langle K^* \vert {\bar s} \{ \gamma^\mu  - 2 v^\mu \} 
(1 - \gamma_5) b \vert B \rangle \big]
\end{eqnarray}
In the rest frame of the $B$-meson $[v^\mu = (1,{\vec 0})]$
 and we have $v \cdot \eta$ = 0 (from the transversality of the photon).
 Thus the term in eq.7 containing
$v^\mu$ is unable to contribute.
 Therefore, 
in the limit of large $b$ and $s$-quark masses 
the hadronic matrix elements of the tensor current get expressed in
terms of the
vector and axial vector form-factors (functions of $y\equiv v\cdot v^\prime$),
which in turn admit the
following general decomposition consistent with Lorentz invariance 
\cite{neubert},
to wit,
\begin{equation}
\langle K^* (v') \vert V_\mu^{sb} \vert \bar B (v) \rangle
= i {\sqrt {M_B M_K^*}} \xi_V (y) \epsilon_{\mu\nu\alpha\beta}
\epsilon^{\nu} v'^{\alpha} v^{\beta}
\end{equation}
\begin{eqnarray}
\langle K^* (v') \vert A_\mu^{sb} \vert \bar B (v) \rangle
 & = &  {\sqrt {M_B M_K^*}}\{ \xi_{A1} (y) (y + 1) \epsilon_\mu^*
                 -\xi_{A2} (y) \epsilon^* \cdot v v_\mu\nonumber\\
& &  -\xi_{A3} (y)
 \epsilon^* \cdot v v_\mu'
					  \}		
\end{eqnarray}
where $\epsilon$ is the polarisation vector of $K^*$,
 $V_\mu^{sb} = {\bar s} \gamma_\mu b$ and $A_\mu^{sb} = {\bar s}
\gamma_\mu \gamma_5 b$. The factor $\sqrt{M_B M_{K^*}}$ represents
the change of normalisation of the states from the usual covariant
one to that appropriate for HQET. From the underlying kinematics  
$y \equiv v \cdot v' = { M_B^2 + M_{K^*}^2 \over 2 M_B M_{K^*}}$. 
In the heavy quark limit it is very convenient to use a matrix
representation of the mesonic states \cite{neu} namely 
$B=-{ 1+ \vslash \over 2}\gamma_5$
and $K^* = {1+\vpslash \over 2} \epslash$, whereupon the matrix
elements of the vector and axial vector currents may be written as
\begin{equation}
 \langle K^* (v') \vert V_\mu^{sb} \vert \bar B (v) \rangle
= i {\sqrt {M_B M_K^*}} \xi (y) Tr\{\epslash{(1+\vpslash)\over 2}\gamma_\mu
 {(1+\vslash)\over 2}\gamma_5\}
\end{equation}
\begin{equation}
\langle K^* (v') \vert A_\mu^{sb} \vert \bar B (v) \rangle
  =   {\sqrt {M_B M_K^*}}\xi (y) Tr\{\epslash{(1+\vpslash)\over 2}
\gamma_\mu \gamma_5
 {(1+\vslash)\over 2}\gamma_5\}
 \end{equation}                
the functions $\xi_i (y)$ of eqs.(11) and (12) get related to a
single universal form factor, $\xi(y)$, the Isgur-Wise function,
through 
\begin{equation}
\xi_i (y) = \alpha_i \xi(y)
\end{equation}
with
$
\alpha_V = \alpha_{A1} = \alpha_{A3} = 1
$
and
$
\alpha_{A2} = 0
$. The Isgur-Wise function
satisfies the condition $\xi (1) = 1$ at zero recoil ($y = 1$). 
Thus in  leading order the decay width is given by
\begin{eqnarray}
\Gamma (B \rightarrow K^* \gamma )
 & = & {G_F^2 \over {64\pi^4}}\alpha \vert V_{tb} \vert^2 
\vert V_{ts} \vert^2 \vert
C_7(\mu)\vert^2 m_b^2   \nonumber\\
 & & (1 -{{M_{K^*}^2}\over {M_B^2}})
{ M_{K*}} \vert \xi (y) \vert^2 (y+1) \nonumber\\
& & [(M_B - M_{K^*})^2 (y+2) + (M_B + M_{K^*})^2 (y-1)]
\end{eqnarray}
 
  Whereas the b-quark may safely be considered to be heavy
 with respect to the QCD energy scale, this is at best an uncertain
assumption in the case of the 
strange quark, though this has been the basis of several estimates \cite{ali}.
To test the efficacy of such calculations it is important to study
the magnitude of the ${1\over m_s}$ correction. This correction
originates from two sources, namely, from the current on the one hand
and the mesonic states on the other. 

Consider the current 
$ {\bar h_b} \Gamma u_s$ (with $\Gamma$ =$\gamma_\alpha$ or
$\gamma_\alpha \gamma_5$). For the $b$-quark one can 
with impunity used the
HQET effective field satisfying the condition $\vslash h_b =
h_b$. As we wish to calculate the non-leading
corrections \cite{luke} in the case of the strange quark, we begin by 
recognising that its momentum is
$m_s v^\prime + l$ (where $l$ is a measure of the off-shellness of
that quark in the daughter meson) and hence one has the equation of motion
\begin{equation}
(m_s \vpslash +\lslash -m_s) u_s = 0
\end{equation}
which yields $(1-\vpslash )u_s
={\lslash \over m_s} u_s$, while in the limit of a heavy
$s$-quark one has $(1-\vpslash )u_s = 0$. Thus the
relevant correction arising from the current is given by
\begin{equation}
{\bar h}_b \Gamma u_s ={\bar h}_b \Gamma ({{1+\vpslash}
\over 2} + {{1-\vpslash }\over 2})u_s
= {\bar h}_b \Gamma (1+{\lslash \over 2 m_s}) h_s + O({1\over
m_s^2})
\end{equation}
Implementing this correction in the matrix element of the vector and
axial vector currents between the relevant hadronic states and  using the
trace formalism for the mesons
we arrive at the structure of the $1/m_s$ correction emanating from
the current
\begin{equation}
-{1\over 2m_s} Tr[ \xi^\alpha \epslash^*({1+\vpslash \over 2})\gamma_\alpha
\Gamma ({1+\vslash\over 2})\gamma_5]
\end{equation}
where $\xi^\alpha$, representing the expectation of $l^\alpha$,
 has the generic form 
\begin{equation}
\xi^\alpha = \xi_+ (v+v^\prime)^\alpha + \xi_- (v-v^\prime)^\alpha +
\xi_3 \gamma^\alpha
\end{equation}
However, the three functions so introduced  are not independent.
Indeed using the equation of motion  $iv.D h_b =0$ 
($D$ being the covariant derivative),
the $\xi_i^{\prime}s$ appearing here get related to each other through
 \begin{equation}
\xi_3 =- (1+y)\xi_+-(1-y)\xi_-
\end{equation}
Again, the relation  $\partial_\mu(\bar {h_s}\Gamma h_b) = \bar {h_s}
{\stackrel{\leftarrow}{D _\mu}}
\Gamma h_b +\bar {h_s}{\stackrel{\rightarrow}{D _\mu}}\Gamma h_b$ when 
sandwiched 
between
meson states yield the condition 
 \begin{equation}
\xi_- = {{\bar \Lambda}\over 2}\xi
\end{equation}
where ${\bar \Lambda}$ is the mass of the light degrees of freedom
including the binding energy.
Thus the $1\over m_s$ correction arising from the modification of the
current introduces only one new function  $ \xi_+$ .

As for the corrections arising from the meson states \cite{neu}, it may be
observed that $\vert M(v)\rangle$ is an eigenstate of the leading
order effective Lagrangian of HQET,
 \begin{equation}
{\cal{L}} = \bar {h_s} iv\cdot D h_s
\end{equation}
While, the physical mesonic state $\vert M(p)\rangle$ is an eigenstate 
 of full QCD. 
Expanding the Lagrangian in powers of ${1\over m_s}$ we may write
\begin{equation}
{\cal L} = {\cal L}_{HQET} +{1\over 2m_s}{\cal L}_1 +{1\over
4m_s^2}{\cal L}_2 +\cdots
\end{equation}
with 
\begin{eqnarray}
{\cal L}_1 & = &  \bar{h_s}(iD)^2 h_s + {g\over 2}\bar{h_s} \sigma^{\mu
\nu} G_{\mu \nu} h_s\nonumber\\
& = & {\cal O}_1 + {\cal O}_2
\end{eqnarray}
 Proceeding perturbatively the  corrections to order ${1\over m_s}$
  is given by 
\begin{equation}
{1\over 2m_s }{\langle}
 K^*\vert i\int dy T\{\bar{h_s}\gamma_{\mu}h_b(0), {\cal L}_1(y)\}\vert
 B\rangle 
\end{equation}
from which the term containing ${\cal O}_1$, namely
\begin{equation}
-{{\bar \Lambda}\over {2 m_s}}\{\psi_1(y)Tr[\epslash{{(1+\vpslash)}
\over 2} \Gamma{{(1+\vslash)}\over 2 }\gamma_5]\}
\end{equation}
 can be estimated, while that pertaining to ${\cal O}_2$ is 
\begin{eqnarray}
& &-{{\bar \Lambda}\over {2 m_s}}
\{i\psi_2(y)Tr[v_{\mu}\gamma_{\nu}
\epslash{{(1+\vpslash)}
\over 2 }\sigma^{\mu\nu}{{(1+\vpslash)}\over 2} \Gamma{{(1+\vslash)}\over 2}
\gamma_5] \nonumber\\
& +&\psi_3(y)Tr[\sigma_{\mu\nu}
\epslash{{(1+\vpslash)}
\over 2 }\sigma^{\mu\nu}{{(1+\vpslash)}\over 2} \Gamma{{(1+\vslash)}\over 2}
\gamma_5]\} 
\end{eqnarray}
Thus three new functions $\psi_1$, $\psi_2$ and $\psi_3$ make their appearence
 from the $1\over m_s$ correction to the meson states.
 From Luke's theorem, which states that the ${1\over m_s}$ correction
must vanish at the point $y=1$, one arrives at the constraints 
 \begin{equation}
 \psi_1(y =1) =\psi_3(y=1) =0
\end{equation}
The four form-factors arising  in the
next-to-leading order approximation 
viz. $\xi_+(y)$, $\psi_1(y)$, $\psi_2(y)$ 
and $\psi_3(y)$
may for convenience be re-expressed in terms of $\rho_i$ ($i = 1~~{\rm
to}~~4$):
\begin{equation}
\rho_1(y)\xi(y) = {{\bar \Lambda}\over 2}[ \psi_1(y) - 2 (y-1)\psi_2(y)
+6\psi_3(y)
\end{equation}
\begin{equation}
\rho_2(y)\xi(y) = {{\bar \Lambda}\over 2}[ \psi_1(y) - 2 \psi_3(y)
\end{equation}
\begin{equation}
\rho_3(y)\xi(y) = {\bar \Lambda} \psi_2(y) 
\end{equation}
\begin{equation}
\rho_4(y)\xi(y) =- {{\bar \Lambda}\over 2}[(1+y) \xi_+(y) +(1-y)\xi(y)]
\end{equation}

 Thus including  
 corrections from both the sources
 (current and state modifications)
 the formfactors $\xi_i$ (defined in eqs. 11 and 12)  
  could be writen as 
   \begin{equation}
  \xi_i = [\alpha_i + \gamma_i ] \xi
   \end{equation}
 \noindent where $\gamma_i$ are given in terms of 
$\rho_i(y)$s by
 \begin{equation}
 \gamma_{V} = {1\over 2} {{\overline \Lambda}\over m_s} +{1\over m_s}
 \rho_2(y)
 \end{equation}

 \begin{equation}
 \gamma_{A_1} = {1\over 2} {{\overline \Lambda}\over m_s}{{y-1}\over
  {y+1}}+{1\over m_s}
 \rho_2(y)
 \end{equation}

  \begin{equation}
  \gamma_{A_2} = {1\over {y+1}} {1\over m_s}[-\overline  \Lambda +
  (y+1)\rho_3(y)-\rho_4(y)]
  \end{equation}
  \begin{equation}
  \gamma_{A_3} = {1\over 2} {{\overline \Lambda}\over m_s}{{y-1}\over
  {y+1}}+{1\over m_s}(\rho_2(y)-\rho_3(y)-{1\over {y+1}}
 \rho_4(y))
 \end{equation}
 Accordingly the decay width for $B\to K^* \gamma $ 
 including $1/m_s$ correction
 is given by,
\begin{eqnarray}
\Gamma (B \rightarrow K^* \gamma )
 & = & {G_F^2 \over {128\pi^4}}\alpha \vert V_{tb} \vert^2 
\vert V_{ts} \vert^2 \vert
C_7(\mu)\vert^2 m_b^2 \vert \xi (y) \vert^2 (1 -{{M_{K^*}^2}\over {M_B}})
{ M_{K*}}\nonumber\\
 & &  [(M_B - M_{K^*})^2 \{3(1+2\gamma_{A_1})(y+1)^2 +(1+2\gamma_{A_3})
 (1-y^2)\}\nonumber\\
& & +2 (M_B + M_{K^*})^2 (1+2\gamma_V)(y^2-1)]
\end{eqnarray}
 However, the form of the Isgur-Wise function (as also those which
arise in the next-to-leading 
 order) are model dependent and different models lead to a range
 of values for the the width being considered. Here we shall confine 
 our attention to such models which provide estimates
 of the non-leading orders,
 for it is the purpose of this work to study this aspect of the problem.

\bigskip

\centerline {\Large \bf III. MODELS AND RESULTS}
	
	The form of the Isgur-Wise function, as also $\rho_i$ (i=1 to 4),
shall next be obtained from the  improved ISGW model \cite{iisgw} as well as
from the BSW  
 model \cite{bsw}.

\centerline{\bf 1. Improved Isgur-Scora-Grinstein-Wise Model}
\vskip .5cm
In the ISGW model \cite{isgw}, an S-wave $Q{\bar d}$ meson $X$ (here $Q = b$ or $s$
 and $X= B$ or $K^*$) is represented by,
\begin{equation}
\vert X(\vec{p_X},{\sl s}_X)\rangle = \int d^3 \vec k \sum_{{\sl s},
\bar {\sl s}}
\phi_X( k)
 \chi_{{\sl s}{\bar s}}\vert Q[\vec p_Q(\vec p_X,\vec  k),s]
{\bar d}[\vec p_d(\vec p_X,\vec k), r]\rangle
\end{equation}
where $\phi_X( k)$ is the wave function of the relevant $Q{\bar d}$
S-wave state as a funtion of the relative momentum k,
 $\chi_{{\sl s}{\bar s}}$ describes 
the coupling of the spins of the constituent
 quarks into the meson (pseudoscalar or vector, having polarisation
vector $\epsilon$). The momenta of the quarks, $\vec p_Q$  and $\vec p_d$,
 are given by,
\begin{equation}
\vec p_Q = m_Q \vec  v_X +\vec k
\end{equation}
\begin{equation}
\vec p_d = m_d\vec v_X-\vec k
\end{equation}
where $\vec  v_X$ is velocity of the meson and $\vec k$ is the exchange 
momentum of the constituent quarks inside
the meson.

The wave function $\phi_X( k)$ is taken as the Fourier transform of
 the solution of the
Schr{\"o}dinger equation with a Hamiltonian 
\begin{equation}
H_{ISGW} = -{\nabla_Q^2\over 2m_Q}  -{\nabla_d^2\over 2m_d}
			-{4\alpha_s \over 3r} + a_1 r +a_2
\end{equation}
and using the variational method with  oscillator wave-functions as
trial functions for
the required ground states viz.
\begin{equation}
\psi_{1S}(r) =({\beta_S\over{\sqrt \pi}})^{3/2} 
				\exp [-\beta_S^2 r^2 /2]
\end{equation}
and, evaluating the matrix element of the current, and
 going to the heavy quark limit ($m_Q =M_X$ and $\phi_B = \phi_{K^*}$),
provides us with
the ISGW estimates for the leading order form-factors, which, while
in accordance with the heavy quark symmetry at the kinematic point
$y=1$,  fails to conform at arbitrary
values of $y$ (as the same universal Isgur-Wise function do not emerge
from all the form-factors). In order to rectify this deficiency 
 Amundson \cite{iisgw} proposed an improved version of the ISGW model where 
the pseudoscalar
(in our case $B$) and vector (here $K^*$) meson  states are given by
\begin{equation}
\vert B(v) \rangle 
= \int d^3 \vec {k} \sum_{{\sl r}, {\sl s}}
\phi_B( {k}){\bar u_{\sl s}}
(p_b(v,k),s)\gamma_5 
 v_d(p_d(v,k),{ r})
\end{equation}
and
\begin{equation}
\vert K^*(v^{\prime}) \rangle 
= \int d^3 \vec {k^{\prime}} \sum_{{\sl r}^{\prime}, {\sl s}^{\prime}}
\phi_K^*( {k^{\prime}}){\bar u_{\sl s}}
(p_s(v^{\prime},k^{\prime}),{ s}^{\prime})\epslash 
v_d(p_d(v^{\prime},k^{\prime}), { r}^{\prime})
\end{equation}
The heavy quark limit is easily arrived at by replacing the meson
wave-functions $\phi_M$ by  
$\phi_{\infty}$ (the limit of $\phi_M$ as $M_Q\rightarrow \infty$),
which is the solution of the Hamiltonian, 
\begin{equation}
H_0 = -{\nabla_d^2\over 2m_d} +V_{spinless}
\end{equation}
where $V_{spinless}$ is the part of the quark-antiquark 
binding potential inside
the meson that does not depend on the spins of the constituent quarks.
Thus the relevant matrix elements are
\begin{eqnarray}
\langle K^{*}\vert V^{\mu}\vert B\rangle
& = & -\int d^3 k^{\prime} d^3 k \phi_{\infty}(k^{\prime})\phi_{\infty}(k)
\delta^3(\vec p_d -\vec{p_d^{\prime}})\nonumber\\
& & \sum_{r r^{\prime} ,s s^{\prime}}
{\bar v_d(p_d^{\prime}(v^{\prime},k^{\prime}),r^{\prime})}\epslash
 u_s
(p_s(v^{\prime},k^{\prime}),s^{\prime})\nonumber \\
& & {\bar u_s}
(p_s(v^{\prime},k^{\prime}),s^{\prime})\gamma^{\mu}
 u_b
(p_b(v,k),s)\nonumber\\
& & {\bar u_b}
(p_b(v,k),s)\gamma_5v_d(p_d(v,k),r)\nonumber\\
& = & -Tr[\epslash {{1+\vpslash}\over 2}\gamma^{\mu}{{1+\vslash}\over 2}
\gamma_5]\nonumber\\
& & \int d^3 k^{\prime} d^3 k \phi_{\infty}(k^{\prime})\phi_{\infty}(k)
\delta^3(\vec p_d -\vec {p_d^{\prime}})
\end{eqnarray}
 and 
\begin{eqnarray}
\langle K^{*}\vert A^{\mu}\vert B\rangle
& = & -\int d^3 k^{\prime} d^3 k \phi_{\infty}(k^{\prime})\phi_{\infty}(k)
\delta^3(\vec p_d -\vec{p_d^{\prime}})\nonumber\\
& &\sum_{r r^{\prime} ,s s^{\prime}}
{\bar v_d(p_d^{\prime}(v^{\prime},k^{\prime}),r^{\prime})}\epslash
 u_s
(p_s(v^{\prime},k^{\prime}),s^{\prime})\nonumber\\
 & & {\bar u_s}
(p_s(v^{\prime},k^{\prime}),s^{\prime})\gamma^{\mu}
   \gamma_5 u_b
(p_b(v,k),s)\nonumber\\
& & {\bar u_b}
(p_b(v,k),s)\gamma_5v_d(p_d(v,k),r)\nonumber\\
& = & -Tr[\epslash {{1+\vpslash}\over 2}\gamma^{\mu}\gamma_5
{{1+\vslash}\over 2}
\gamma_5]\nonumber\\
& & \int d^3 k^{\prime} d^3 k \phi_{\infty}(k^\prime)\phi_{\infty}(k)
\delta^{(3)}(\vec p_d -\vec{p_d^{\prime}})
\end{eqnarray}
From the above equations, the universal Isgur-Wise function is
identified as,
\begin{equation}
\xi_(y) = \int d^3 k^{\prime} d^3 k \phi_{\infty}( {k^{\prime}})
\phi_{\infty}( k)
\delta^{(3)}(\vec p_d -\vec p_d^{\prime})
\end{equation}
which after performing the integrations is given by,
\begin{equation}
\xi(y) = \exp[-{m_d^2\over {2\beta^2}}(y-1)]
\end{equation}
The deficiency in the nonrelativistic model is compensated for by
introducing 
a phenomenological parameter $\kappa$ modifying the exponent in $\xi$
replacing $\beta$ by $\kappa \beta$. 
The value of $\kappa $ is fixed to be $0.6\pm .06$ 
 by fitting with the experimentally determined value of the ``charge
radius" ($\rho = 0.93 \pm 0.10$) corresponding to the usual parametrization
of the Isgur-Wise function   $\xi(y)$
as $ 1 - \rho^2 (y-1) + {\cal O}[(y-1)^2]$.

Similarly, the current correction gives a contribution 
\begin{equation}
 -{1\over {2m_s}}Tr[A_{\alpha}\epslash {{1+\vpslash}\over 2}\gamma^{\alpha}
\Gamma^{\mu}{{1+\vslash}\over 2}
\gamma_5]
\end{equation}
where
\begin{equation}
A_\alpha=
\int d^3 k^{\prime}\int d^3 k \phi_{\infty}(\vec {k^{\prime}})
\phi_{\infty}(\vec k)l_\alpha (k^\prime)
\delta^{(3)}(\vec p_d -\vec p_d^{\prime})
\end{equation}
Comparing the results  of the  improved ISGW model ( eq.(52) and eq.(53) )
 with the general expression for the current correction (eq.(19) - eq.(22)),
 we get the two unknown parameters, arising from the current correction
 of order $1\over m_s$, namely
 \begin{equation}
\bar {\Lambda} = m_d
\ee
and
\be
\xi_+(y) = 0
\ee
It may be observed that in this model 
 the term of the form $(1-y)\xi\gamma_{\alpha}$, expected in general,
is absent; this is a 
 limitation of this nonrelativistic model.

Now the correction to the wave function from the 
 heavy quark limit $\phi_{\infty}$
 has to be taken into account.
 Incorporating this correction, the wavefunction has  the form,
 \be
\phi_{K^*} = \phi_{\infty} +\phi^1
\ee
where $\phi^1$ is the correction to be estimated 
in the first order of perturbation.
 This introduces a contribution to the matrix element of the vector and 
axial vector current involving
\be
\int d^3 k^{\prime}\int d^3 k \phi^1(\vec {k^{\prime}})
\phi_{\infty}(\vec k)
\delta^{(3)}(\vec p_d -\vec p_d^{\prime})
\ee
 To calculate these corrections \cite{vera}, one takes the  kinetic energy term 
$({\cal H}_{KE})$ of 
heavy quark $
 -{\nabla_d^2\over 2m_s}$ 
 and the spin-spin interaction   $({\cal H}_{SS})$ term
 ${{gS_d.S_s} \over {2m_d m_s}}\delta^{(3)}(\vec 
 r)$
 as perturbation,  
 to obtain 
 \be
\phi_{K^*}= \phi_{\infty}+\phi^1_{KE} +\phi^1_{SS}
\ee
where, denoting the nth wavefunction of the complete set of basis states 
 (the eigenstates of the unperturbed Hamiltonian) by $\phi_n$,
\bea
\phi^1_{KE} & = & \sum_{n\not = \infty}{{\phi_n(r)}\over{E_n -E_\infty}}
\int d^3{r^\prime}\phi_n^*({r^\prime}){\cal H}_{KE}\phi_{\infty}({r^\prime})
\nonumber\\
& \approx & -[{3\over 2}]^{1/2}{{\beta^2}\over 
 {2m_s(E_{2S}-E_{1S})}}\psi_{2S}(r)
\eea
retaining only the 2S-state expected to give a significant
contribution with  
\begin{equation}
\psi_{2S}(r) = \sqrt{{2\over 3}} 
 				({\beta_S\over{\sqrt \pi}})^{3/2} 
				(\beta^2 r^2 - {3\over 2}) 
\exp [-\beta_S^2 r^2 /2]
\end{equation}

Comparing  equation(57) with equation(27) and inserting the model
wavefunctions from from eq.(44) and eq.(59), we obtain 
 \bea
\psi_1 & = & {{2 m_s}\over {\bar \Lambda}} 
\int d^3 k^{\prime}\int d^3 k \phi^1_{KE}({k^{\prime}})
\phi_{\infty}( k)
\delta^{(3)}(\vec p_d -\vec p_d^{\prime})\nonumber\\
& = & {m_s\over {2(E_{2S} - E_{1S})}}(y-1)\xi(y)
\eea
The energy denominator 
\be
 E_{2S} -E_{1S} = {\beta_S ^2\over m_s} + {a_1\over {\beta_S\sqrt{\pi}}}+
{{4\alpha_s\beta_S}\over {9\sqrt{\pi}}}
\ee
has the value   829 MeV  with $\beta_S$ and $a_1$ taken to[ref] be
0.42 and  0.18Ge$V^2$ respectively.
The observed \cite{ahmady} mass splitting of the 1S 
 and 2S resonant states  of $K^*$ is about 818 MeV.
 The discrepancy is 
less than $1.5\%$ which is acceptable considering the non-relativistic nature 
 of the model.
 
Similarly,  for the  hyperfine 
 interaction term ,
\bea
\phi_{SS}^1 & = &  \langle S_d.S_s\rangle
\sum_{n {\not =}\infty}{\phi_n(r)\over {E_n -E_{\infty}}}
\int d^3 {r ^\prime } \phi_n^*(r ^\prime)
  {{g \delta(r^\prime)}\over {2m_d m_s}}
\phi_{\infty}(r^{\prime})\nonumber\\
& \approx & {1\over 4} {g\over {2m_d m_s (E_{2S} - E_{1S})}}\psi^{2S*}(0)
\psi^{1S}(0)
\psi^{2S}(r)
\eea
with the dominant contribution coming  from the first radial excitation 
$\psi^{2S}$.

 In this model, as the  variational method is used for the unperturbed 
wavefunction, it is not very reliable for the estimation of the value of the
 wavefunction at the origin
 which, however,  can be estimated 
 from available experimental data as
\be
M_{K^*} -M_K = {g\over {2m_d m_s}}\vert \psi^{1S}(0)\vert^2
\ee
For potential models noting that  
\be 
{d^2\over {dr^2}} V(r)\gtlt 0 \Rightarrow
\vert \psi^{2S}(0)\vert^2
\gtlt\vert \psi^{1S}(0)\vert^2
\ee
we may use the fact that for heavy-light systems,
 the linear part of the potential is dominant for which 
${d^2\over {dr^2}}V(r)\approx 0$
  we are led to the interesting result
\be
\psi^{2S}(0) \approx \psi^{1S}(0)
\ee
Accordingly we can rewrite $\phi_{SS}^1$ as,
 \be
{1\over 4}{{M_{K^*} -M_K}\over {E_{2S} - E_{1S}}}\psi^{2S(r)}
\ee
 and again comparing eq. (57) with this form of $\phi_{SS}^1$ and eq.(28)
and inserting the relevant wave-functions, we have
\bea
\psi_3(y) & = & - {m_s\over {\bar \Lambda}} \int d^3k^{\prime}d^3k\phi^1_{SS}
(
{k^\prime})\phi_{\infty} ( k)\delta^3(\vec p_d -\vec {p_d^\prime})
\nonumber\\
& = & {{m_s m_d (M_{K^*}-M_k)}\over {4 \beta^2{\sqrt 6}(E_{2S} -E_{1S})}}
(
y-1)\xi(y)
\eea
Thus the flavor and spin symmetry breaking $\psi_1$ and $\psi_3$ can
be calculated.  
But in this model, there is no symmetry breaking term which can give
a non-vanishing  
 $\psi_2$ (here $\psi_2(y) =0$).
Accordingly, implementing the above results the decay width 
 turns out to be $1.1\times 10^{-17} $GeV
in the heavy quark limit of b and s quark with $\beta_S = 0.42$.
 This, however, is 
  increased to $1.8\times 10^{-17}$ GeV when $1/m_s$ corrections
 are taken into account.
\vskip .5cm
\centerline{\bf 2. Bauer-Stech-Wirbel Model}
\vskip .5cm
The BSW model begins with the definition of the hadronic form-factors 
through
\begin{eqnarray}
\langle K^* \vert j_\mu \vert B \rangle
 & = &
{2\over M_B+M_K^*} \epsilon_{\mu\nu\rho\sigma} \epsilon^{*\nu} p^\rho
p^{\prime \sigma} V(q^2)
\nonumber\\
 & & +i\{ \epsilon^*_\mu (M_B+M_{K^*}) A_1(q^2)
-{\epsilon^*\cdot q\over M_B+M_{K^*}}(p+p^\prime)_\mu A_2(q^2)
\nonumber\\
 & & -{\epsilon^*\cdot q\over q^2} 2M_{K^*} q_\mu A_3(q^2)\}  
   + i {\epsilon^*\cdot q\over q^2} 2M_{K^*} q_\mu A_0(q^2)\}  
\end{eqnarray}
where $A_3(q^2)$ is simply an abbreviation for
\begin{equation}
A_3(q^2) = {M_B+M_{K^*} \over 2M_{K^*}} A_1(q^2) - {M_B-M_{K^*}\over
2M_{K^*}} A_2(q^2)
\end{equation}
and in order to cancel the artificially introduced 
singularity at $q^2 = 0$ we must have
$A_3(q^2=0) = A_0(q^2=0)$. One goes on to relate these form-factors
 at maximum recoil (which is all that is needed here because for the real
photon we have $q^2=0$) 
 to the overlap integral of the initial and final mesonic
wave-functions in the infinite momemtum frame, which depends on the
fraction ($x$) of the longitudinal momentum carried by the heavy
quark and its square-averaged transverse momentum ($\omega^2 =
\langle p_T^2 \rangle$), and thus
\begin{eqnarray}
V(q^2 = 0) & = &  {{m_b -m_s}\over {M_B- M_{K^*}}} \langle K^*\vert
{1\over x}\vert B\rangle\nonumber\\
& = &   {{m_b -m_s}\over {M_B- M_{K^*}}}\int_0^1 dx
\phi_{K^*}(x){1\over x}\phi_B(x)
\end{eqnarray}
\begin{eqnarray}
A_1(q^2 = 0) & = &  {{m_b +m_s}\over {M_B+ M_{K^*}}} \langle K^*\vert
{1\over x}\vert B\rangle\nonumber\\
& = &   {{m_b +m_s}\over {M_B+ M_{K^*}}}\int_0^1 dx
\phi_{K^*}(x){1\over x}\phi_B(x)
\end{eqnarray}
\begin{eqnarray}
A_3(q^2 = 0) & = &   \langle K^*\vert
 B\rangle\nonumber\\
& = &   \int_0^1 dx
\phi_{K^*}(x)\phi_B(x)
\end{eqnarray}
Here $\phi_{K^*}(x)$ and $\phi_B(x)$  are taken  generically from
 the ground state
solution of a relativistic scalar harmonic oscillator potential, i.e.
 \begin{equation}
\phi_M(x)= N_M \sqrt{x(1-x)}\exp{[-{M^2\over {2\omega^2}}(1 -x -
{\alpha\over M})^2]}
\end{equation}
 where $M = M_B$ or  $M_{K^*}$. Taking the heavy quark limit one
arrives at the Isgur-Wise function in the leading approximation, to wit,
\begin{equation}
\xi_{BSW}(y) = {2\over y+1} \exp [-{2m_d^2 \over \omega^2}({y-1\over y+1})]
\end{equation}
and  the functions appearing in the next 
to leading order ($1\over m_s$ corections) are given by
\begin{equation}
\rho_1^{BSW}(y) ={\Delta (y) \over y+1}
\end{equation}
\begin{equation}
\rho_2^{BSW}(y) = -{y-\sqrt{y^2-1}\over y+1} \Delta (y)
\end{equation}
\begin{equation}
\rho_3^{BSW}(y) = (1-\sqrt{y-1\over y+1}) {\Delta (y)\over y-1}
\end{equation}
\begin{equation}
\rho_4^{BSW}(y) = 0
\end{equation}
where
\begin{equation}
\Delta (y) = {y-1\over 2y}\alpha + {\omega \over 2} 
[I({\alpha\over\omega}) -\sqrt{{y+1\over 2y}} I(\sqrt{{y+1\over 2y}}
{\alpha\over \omega})]
\end{equation}
$\alpha$ being the difference between the values of the masses of the
meson and that of the heavy quark contained therein
and $\bar \Lambda$ as defined in eq.(22) is given in this model by
\begin{equation}
\alpha [ 1+{\omega\over 2\alpha} I({\alpha \over \omega})]
\end{equation}
In all these expressions the function $I$, related to the error
integral, is given by
\begin{equation}
I(x) \equiv {\int^\infty_{-x} dz \exp (-z^2)
					\over \int^\infty_{-x} dz (z+x) \exp (-z^2)}
\end{equation}
In this BSW model, one takes the value[ref] $\omega = 400$MeV irrespective
 of the flavor involved, though of course the value of the parameter $\alpha$,
 which is related to the mass difference between the heavy meson and
the heavy quark would depend, in general, on the meson being
considered and the flavor occuring therein. However, in the heavy
quark limit (for both b and s) one should take this parameter to be
flavor independent and we adopt the value[ref] $\alpha = 280$ MeV.
 With these values of
the parameters, the decay width 
 for $B\to K^* \gamma$ (both b and s taken  as heavy) comes out to be 
 $ 2.2\times 10^{-17}$
 GeV, wherein  after incorporating the $1/m_s$ correction 
 one arrives at the value   
 $3.8\times 10^{-17}$ GeV.
\vskip .5cm

\centerline{\bf 3. A general treatment assuming only the b-quark as heavy}
\vskip .5cm
Lastly, we consider the most reliable assumption,
  where only the bottom quark mass 
is taken to be infinity. The hadronic matrix element of tensor-type current
 relevant  for $B\to K^*\gamma$ may be represented as \cite{iw},
\bea
\langle K^*(p^\prime ,\epsilon)\vert {\bar s} \sigma_{\mu \nu}b\vert
 B(p)\rangle & = & g_+\epsilon_{\mu\nu\lambda\sigma}\epsilon^{*\lambda}
(p+p^\prime)^\sigma +
  g_-\epsilon_{\mu\nu\lambda\sigma}\epsilon^{*\lambda}
(p-p^\prime)^\sigma \nonumber\\
& + &
  h\epsilon_{\mu\nu\lambda\sigma}(p+p^\prime)^\lambda
(p-p^\prime)^\sigma (\epsilon^* \cdot p)
\eea
using the relation 
$\sigma^{\mu\nu} = {i\over 2}\epsilon^{\mu\nu\lambda\sigma}
\sigma_{\lambda \sigma}
\gamma_5$, in the preceeding equation, we may write
\bea
\langle K^*(p^\prime ,\epsilon)\vert {\bar s} \sigma_{\mu \nu}\gamma_5 b\vert
 B(p)\rangle & = & ig_+[\epsilon^*_\nu (p+p^\prime)_\mu -\epsilon^*_\mu
(p+p^\prime)_\nu]\nonumber\\
&  + & 
  ig_-[\epsilon^*_\nu (p-p^\prime)_\mu -\epsilon^*_\mu
(p-p^\prime)_\nu]\nonumber\\
&  + & ih[(p+p^\prime)_\nu (p-p^\prime)_\mu \nonumber\\
& & -(p-p^\prime)_\nu
(p+p^\prime)_\mu](\epsilon^*\cdot p)
\eea
In the  heavy quark symmetry limit, as $m_b{\rightarrow}\infty$
$$ \vslash b = b$$
 which in the rest frame of the B meson becomes
\be
\gamma_0 b = b
\ee
Using the above equation the tensor form factors $g_+$ , 
$g_-$ and h get related 
to the vector and axial 
 vector type form factors  V, $A_1$ $A_2$ and $A_3$ through
\be
\langle K^*(p^\prime ,\epsilon)\vert {\bar s} \sigma_{0 i} b\vert
 B(0)\rangle = -i 
\langle K^*(p^\prime ,\epsilon)\vert {\bar s} \gamma_ i b\vert
 B(0)\rangle 
\ee 
\be
\langle K^*(p^\prime ,\epsilon)\vert {\bar s} \sigma_{0 i} \gamma_5 b\vert
 B(0)\rangle = i 
\langle K^*(p^\prime ,\epsilon)\vert {\bar s} \gamma_ i \gamma_5 b\vert
 B(0)\rangle 
\ee
The decay width for $B\to K^*\gamma$ takes the form,
\bea 
\Gamma 
  &  = & {5 \over 4}{G_F^2 \over {128\pi^4M_B^5}}\alpha
 \vert V_{tb} \vert^2 
\vert V_{ts} \vert^2 \vert
C_7(\mu)\vert^2 m_b^2\nonumber\\   
& &   (M_B^2 -M_{K^*}^2)^3\{(M_{K^*}+M_B)A_1(0) +(M_B - M_{K^*})V(0)\}^2
\eea
 $A_1(0)$ and $V(0)$ may then be obtained from the BSW model \cite{bsw2} and
 gives the decay width of magnitude $2.4\times 10^{-17}$GeV.
  The ISGW model is not suitable
 for providing estimates in this approach due to its nonrelativistic nature.
\vskip 1cm
\noindent
To calculate $R(B\to K^* \gamma)$ (the ratio of the decay width of 
 $B\to K^* \gamma$ to that of the quark level process $b\to s\gamma$ 
\cite{grin,grig} 
($B\to X_s\gamma$)) one has to consider the decay width of $b\to s \gamma$,
 expressed as,
\be
\Gamma(b\to s \gamma) = {{G_F^2\alpha}\over {32\pi^4}}
 \vert V_{tb} \vert^2 
\vert V_{ts} \vert^2 \vert
C_7(\mu)\vert^2 m_b^5   
\ee 
The numerical value for the decay width of this quark-level process
is $1.3\times10^{-16} $GeV.

The hadronisation ratio $R=\Gamma(B\to K^* \gamma)/\Gamma(b\to s \gamma)$
 calculated in all these approaches is shown in Table 1.

\vskip .5cm
\centerline{\Large \bf IV. SUMMARY AND CONCLUSION}
\vskip .5cm
We have shown the results for hadronization fraction 
 $R=\Gamma(B\to K^* \gamma)/\Gamma(b\to s \gamma)$, calculated in the
frame work of Heavy Quark Effective Theory. the required formfactors are
 obtained from the improved ISGW as well as from the BSW model, for
both the cases, results are shown first for a heavy s quark ($m_s\sim
\infty$) and then implementing the $1/m_s$ correction. It may be seen
 that in the former case $1/m_s$ correction is as much as $65\%$
while in the latter case it is even larger ($71\%$). Comparing with
the experimentally observed value of $R =0.19\pm 0.09$ \cite{browder}, we see 
 that for the the BSW model the calculated value of R overshoots from
0.17( both b and s heavy) to 0.30 (incorporating $1/m_s$ corrections).
 The fact that taking only b to be heavy and keeping $m_s$ at its
physical value resulting in $R=0.18$ suggest that the higher order
$1/m_s$ corrections contribute with opposite sign.
 In any case, the results strongly indicate that
the approach where the s quark is taken to be heavy is not only
model-dependent but is also highly unreliable as the perturbation 
 expansion in power of $1/m_s$ does not seem to converge well and
alternates in sign.

\newpage
\begin{center}
\vskip .5cm
TABLE I\\
\vskip .5cm
\begin{tabular}{|c|c|c|c|c|}\hline
& & & & \\
Model & Condition & Decay & Branching & R($B\to K^*\gamma$)\\ 
& & width & fraction &\\
 & & in GeV & $={{\Gamma(B\to K^*\gamma)}\over {\Gamma(B\to all)}}$ & 
 $ ={{ \Gamma(B\to K^*\gamma)}\over {\Gamma(b\to s \gamma)}}$\\
& & & & \\ \hline
& & & & \\

Improved  &  b \& s as heavy & 
$1.10$ &  
  $2.51$ & $.09$\\
 ISGW & (without $1/m_s$ &$\times 10^{-17}$  & $\times 10^{-5}$&\\
 &  corrections)&  & &\\
& & & & \\ \hline
& & & & \\

Improved  &  b \& s as heavy & 
$1.82$ &  
  $4.15$ & $.14$\\
ISGW & (with $1/m_s$ & $\times 10^{-17}$  &$\times 10^{-5}$ &\\
 &  corrections)&  & &\\
& & & & \\ \hline
& & & & \\

BSW &  b \& s as heavy & $2.23$ & 
  $5.08$ & $.17$\\
 & (without $1/m_s$ &$\times 10^{-17}$  &$ \times 10^{-5}$ &\\
 &  corrections)&  & &\\
& & & & \\ \hline
& & & & \\

BSW &  b \& s as heavy  & $3.82$ & 
  $8.70$ & $.30$\\
 & (with $1/m_s$ &$\times 10^{-17}$  &$\times 10^{-5}$ &\\
 &  corrections)&  & &\\
& & & & \\ \hline
& & & & \\

BSW &  Only b as heavy & $2.35$ & 
  $5.35$ & $.18$\\
 & &$\times 10^{-17}$  &$\times 10^{-5}$ &\\
& & & & \\
& & & & \\ \hline
\end{tabular} \\
\end{center}
\vskip .5cm
\end{document}